\documentclass[%
 aps,
 amsmath,amssymb,
 reprint,%
superscriptaddress,
]{revtex4-1}

\usepackage{graphicx}
\usepackage{dcolumn}
\usepackage{bm}

\usepackage[utf8]{inputenc}
\usepackage[T1]{fontenc}
\usepackage{mathptmx}
\usepackage{etoolbox}

\usepackage{hyperref}

\makeatletter
\def\@email#1#2{%
 \endgroup
 \patchcmd{\titleblock@produce}
  {\frontmatter@RRAPformat}
  {\frontmatter@RRAPformat{\produce@RRAP{*#1\href{mailto:#2}{#2}}}\frontmatter@RRAPformat}
  {}{}
}%
\makeatother
\begin{document}

\preprint{AIP/123-QED}

\title[Spin Hall effect in a high-resistivity high-entropy alloy]{Spin Hall effect in the high-resistivity high-entropy alloy AlCrMoW}


\author{Jyoti Yadav}%
\affiliation{ 
University Lille, Institute of Electronics, Microelectronics and Nanotechnology, 59000 Lille, France
}%
\affiliation{ 
New Materials Electronics Group, Department of Electrical Engineering and Information Technology, Technical University of Darmstadt, Merckstr. 25, 64283 Darmstadt, Germany
}%
\author{Felix Janus}
\affiliation{ 
New Materials Electronics Group, Department of Electrical Engineering and Information Technology, Technical University of Darmstadt, Merckstr. 25, 64283 Darmstadt, Germany
}%
\author{Tiago de Oliveira Schneider}
\affiliation{ 
New Materials Electronics Group, Department of Electrical Engineering and Information Technology, Technical University of Darmstadt, Merckstr. 25, 64283 Darmstadt, Germany
}%
\author{Shalini Sharma}
\affiliation{ 
Advanced Thin Film Technology Group, Institute of Materials Science, Peter-Grünberg-Straße 2, 64287 Darmstadt, Germany
}%
\author{Daniel Schröter}
\affiliation{ 
New Materials Electronics Group, Department of Electrical Engineering and Information Technology, Technical University of Darmstadt, Merckstr. 25, 64283 Darmstadt, Germany
}%
\author{Markus Meinert}
 \email{markus.meinert@tu-darmstadt.de}
\affiliation{ 
New Materials Electronics Group, Department of Electrical Engineering and Information Technology, Technical University of Darmstadt, Merckstr. 25, 64283 Darmstadt, Germany
}


\date{\today}

\begin{abstract}
We study thin films of the high-entropy alloy system Al$_{x}$(CrMoW)$_{1-x}$, grown on Ta seed layers by magnetron co-sputtering. Between $x=0.2$ and $x=0.6$, a resistivity larger than 100$\mu\Omega$cm is achieved, with a peak of 180$\mu\Omega$cm at $x=0.5$. Around the stoichiometric composition AlCrMoW, the alloy forms a bcc solid solution. The harmonic Hall method was used to characterize the spin Hall angle of the alloy series, where a maximum spin Hall angle of $\theta = -0.12 \pm 0.01$ is observed for $x=0.25$.  The implied spin Hall conductivity is $\sigma_\mathrm{SH} \approx -72\,000 \, \hbar/(2e)$\,S/m. The experimental results show excellent agreement with density functional theory calculations, which show similar trends and values. The results demonstrate that high-entropy alloys with a main-group element component can form a simple crystal structure and show high resistivity. This suggests that a whole new class of materials for spin Hall device engineering is available with simple methods.
\end{abstract}

\maketitle

\section{Introduction}

The spin Hall effect (SHE) converts a charge current density $j$ into a transverse spin current density $j_\mathrm{s}$, with the spin Hall angle (SHA) $\theta = j_\mathrm{s} / j$.\cite{Jungwirth2012, Hoffmann2013}  In cubic, concentrated alloys and pure elements, the SHA is determined by the intrinsic contribution, which is derived from the band structure of the material.\cite{Kdderitzsch2015} It can be written as $\theta_\mathrm{SH} = \sigma_\mathrm{SH} \rho$, where $\sigma_\mathrm{SH}$ is the spin Hall conductivity (SHC) and $\rho$ is the resistivity. Thus, a material with a large SHA requires a large SHC and high resistivity. The two elements with the highest SHC are Pt (positive SHC) and $\beta$-W (negative SHC).\cite{Sagasta2016, Qiao2018, McHugh2020} While Pt crystallizes with the fcc structure and thin films thereof exhibit resistivities below $50\,\mu\Omega$cm, the $\beta$-phase of W crystallizes with the A15 structure and shows resistivities well above $100\,\mu\Omega$cm.\cite{Sagasta2016, Nguyen2016, Dutta2017, Hao2015, Kim2020} Therefore, $\beta$-W routinely shows large SHA values in experiments. However, the $\beta$-phase is metastable and depends critically on the growth conditions and the built-in defects (such as trapped oxygen or nitrogen).\cite{Demasius2016, Neumann2016, Sethu2021, Gmez2023} The stable bcc phase $\alpha$-W has both much smaller SHC and smaller resistivity.\cite{McHugh2020} However, bcc-W still has the largest SHC among all stable allotropes of the elements, with the exception of Pt. Alloying of W and related elements has been proposed to optimize the Fermi level and maximize the SHA.\cite{Sui2017, Kim2020} It was shown that the Fermi level tuning of the SHC is possible with Ta-Re alloys, reproducing the SHC of W at intermediate stoichiometries.\cite{Janus2025a} The resistivity tuning by alloying is a straightforward method to enhance the resistivity and, thereby, the spin Hall angle; this was predicted for multiple Pt alloys and demonstrated experimentally.\cite{Obstbaum2016, Meinert2020, McHugh2024, Janus2025b}

In the present study, we explore the spin Hall effect of a high-resistivity high-entropy alloy (HEA) based on W and the other elements of the same group. A high-entropy alloy forms a simple bcc or fcc crystal structure, which is mainly stabilized by the configurational entropy rather than formation enthalpy.\cite{Miracle2017, Rittiruam2022} Typically materials with at least four elements are referred to as HEA. We refer the reader to the Appendix for a more complete description of what characterizes a HEA in detail. Only few reports on the SHE of HEAs have been published. An early report characterizes a Ta-Nb-Hf-Zr-Ti HEA, for which an SHA of $\left|\theta\right| \approx 0.033$ is reported.\cite{Chen2017} However, in this report the material is amorphous and thereby yields a high resistivity. A recent study investigates Mn-Nb-Mo-Ta-W alloys, which are found to grow as a single bcc phase. Spin Hall magnetoresistance analysis indicates a spin Hall angle of $\left|\theta\right| = 0.12$.\cite{Kubota2025} Another recent study shows a very large SHA in IrRuAlGa-based HEAs with a partially ordered tetragonal structure grown epitaxially on MgO. SHA larger than $\left|\theta\right| = 0.6$ and huge resistivities of more than 200\,$\mu\Omega$cm are reported for multiple compositions.\cite{Wang2025} According to the findings of a recent theoretical high-throughput calculation study, HEAs with a main group element should exhibit high resistivity, in excess of 100\,$\mu\Omega$cm.\cite{Fukushima2022} To form a bcc HEA with a large SHC and large resistivity, we start from an equiatomic mixture of W, Mo, and Cr. The same-group elements Mo and Cr have lower SHC, but still the largest within their respective periods among the bcc elements.\cite{Go2024} By screening the main group elements for suitable values of $\Delta H_\mathrm{mix}$, $\Omega$, and $\delta$ (see Appendix for the definitions), we find that Al is a suitable candidate for HEA formation. While the binary Cr-W and Cr-Mo phase diagrams show miscibility gaps because of the atomic size-mismatch, the other binary phase diagrams contain numerous line compounds and W-Mo forms a solid solution for all compositions. We obtain $\Delta H_\mathrm{mix} \approx -8$\,kJ/mol, $\Omega \approx 3.32$, and $\delta \approx 4.3\%$.  The VEC is 5.25, so formation of a bcc solid solution, similar to the elemental structures of Cr, Mo, and W, is expected.

\section{Methods}

Thin films of the type Si / SiO$_2$ 200nm / Ta 3nm / Al$_{x}$(CrMoW)$_{1-x}$ 8nm / Co$_{40}$Fe$_{40}$B$_{20}$ 3nm / Ta 0.5nm / TaOx 1nm were grown by room-temperature DC magnetron (co-)sputtering with varying Al concentrations. Typical power levels were between 50W and 70W. The individual elemental particle fluxes were determined with a calibrated film thickness sensor. The total growth rate was typically 0.3\,nm/s and the Ar working pressure was $2\times 10^{-3}$\,mbar. We measured the film resistivity using a standard four-point technique as $\rho_\mathrm{eff} = k R_\mathrm{4w} t_\mathrm{total}$, a numerically determined correction factor $k$, and a parallel conductor model to obtain the alloy resistivity. X-ray reflectivity confirmed the thicknesses and the crystallization was monitored by x-ray diffraction (XRD) with a rotating anode diffractometer with Cu anode. Afterwards the samples were laterally patterned using photolithography, ion beam etching, and lift-off lithography for contact lines. The spin Hall angle and spin Hall conductivity were determined with the harmonic Hall technique with magnetic fields up to 1\,T as described in our previous work.\cite{Fritz2018} We explicitly take the anomalous Nernst effect into account and correct for it. The parameters are obtained as averages over at least five Hall crosses per composition.

Additionally, ferromagnetic resonance (FMR) was used to determine the effective magnetization $M_\mathrm{eff}$ and the Gilbert damping parameter $\alpha$. We used our \textsc{OpenFMR} broadband FMR system\cite{deOliveiraSchneider2025} for the detection of the FMR.

We computed the electrical resistivity, spin Hall conductivity, and the spin Hall angle of the Al$_{x}$(CrMoW)$_{1-x}$ (ACMW) series using the SPR-KKR density functional theory code. All calculations were performed at lattice constants given by Vegard's law, and Debye temperatures for SHA calculations were similarly interpolated between the elemental values. The SHA and resistivity calculations were performed for $T = 300\,$K. Details of our methods are given elsewhere.\cite{Fritz2018, Janus2025a, Meinert2020}

\begin{figure}[t!]
\includegraphics[width=8.6cm]{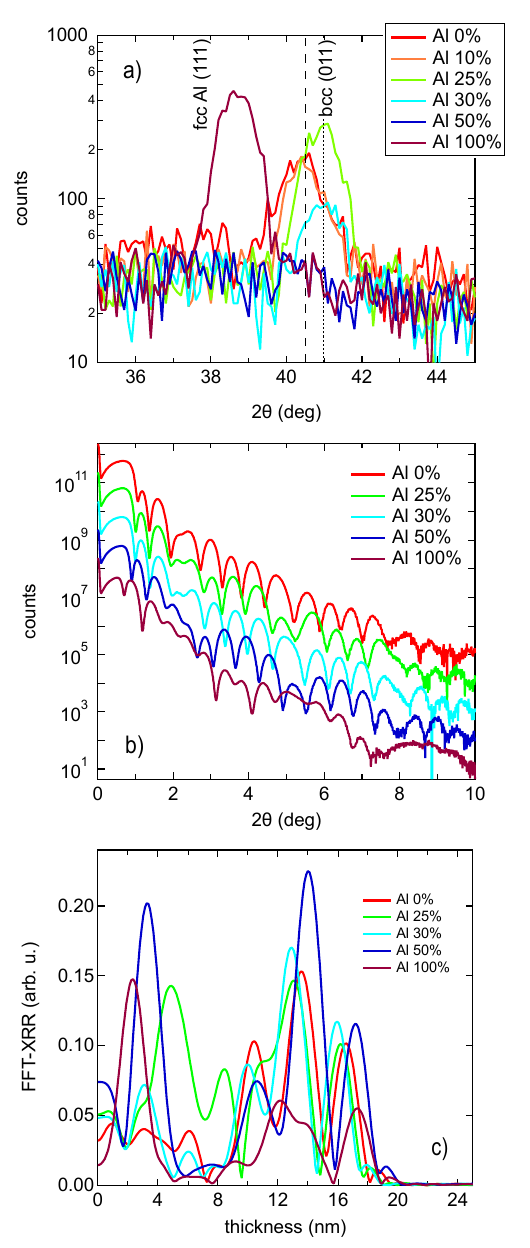}
\caption{\label{fig:XRD}
a) X-ray diffraction measurements of selected samples with Cu K$_\alpha$ radiation. b) X-ray reflectivity measurements on the same samples. The measurements are multiplied by 10 with respect to one another for clarity. c) Fourier transform spectra of the XRR measurements.}
\end{figure}

\section{results}

In Figure \ref{fig:XRD}\,a) we show the XRD measurements of selected samples. We observe a textured growth with a (011) preferred orientation, as expected for a bcc material. A reduction of the lattice constant with increasing Al content is observed and we find the maximum diffraction intensity at the equiatomic composition, i.e. Al 25\%. At Al 50\% and above, no crystallization is observed, with the exception of pure Al, which shows a strong (111) texture. The lattice constant at the equiatomic composition is $a = 3.11$\,\AA{} (dotted line in Fig. \ref{fig:XRD}\,a), very close to the expected value of $a = 4\bar{R} / \sqrt{3} = 3.12$\,\AA{} by simply taking the average atomic radius $\bar{R}$. For the CrMoW alloy (0\% Al) and for a 10\% Al content, we find $a = 3.146$\,\AA{} (dashed line in Fig. \ref{fig:XRD}\,a), which is much larger than the expected value of $a = 3.063$\,\AA{}. However, noting the miscibility gap in both the Cr-W and Cr-Mo phase diagrams, a decomposition into Cr + (Mo,W) seems likely here, and we observe only the bcc solid solution of (Mo,W) with an expected lattice constant of $a = 3.156$\,\AA{}. Further support for this interpretation comes from the increase of the diffraction intensity: the addition of Al to a homogenous solid solution would reduce the diffraction intensity by approximately 30\% due to its small atomic form factor; the actually increased diffraction intensity indicates larger overall crystalline volume and/or improved texture and can be interpreted as due to the formation of a homogenous solid solution. Thus, the entropy effect of the addition of another element to the alloy does indeed stabilize the solid solution phase and the material is correctly characterized as a high entropy alloy. 

All samples showed smooth growth according to x-ray reflectivity (XRR) analysis (Fig. \ref{fig:XRD}\,b)). Typically, the stack roughness parameters of the Gaussian roughness model are $\sigma \approx 3.2$\,\AA{}, where the individual layers are hardly distinguishable. Fitting of multilayer models is notoriously difficult; as an alternative, we present Fourier transform spectra in Fig. \ref{fig:XRD}\,c), which we calculate as described in Ref. \cite{Sakurai2008}. The procedure is described in the Appendix. In the Fourier transform, additional sum and difference peaks appear similar to a beating pattern. The total thickness of all samples is between 16\,nm and 17\,nm, in agreement with the sum of the individual, nominal thicknesses. The most pronounced peak around 13\,nm to 14\,nm arises probably from the sum of the ACMW alloy, the CoFeB, and the cap layer, which all have a similar density. As the magnitude of the oscillation is related to both the roughness of the films as well as to the density contrast at their interfaces, one can not directly relate the Fourier transform magnitudes to the individual roughnesses. One exception is, however, the Al 100\% film, which shows only small oscillation magnitudes and has large roughness. The chosen Ta layer is thus a suitable seed layer to promote crystallinity and smooth growth across the compositional range. In contrast, we observed that growth directly on a SiO$_2$ surface and, alternatively, on a Cr seed layer, led to amorphous films with large roughness at high Al content due to pronounced island growth. 

\begin{figure}[t!]
\includegraphics[width=8.6cm]{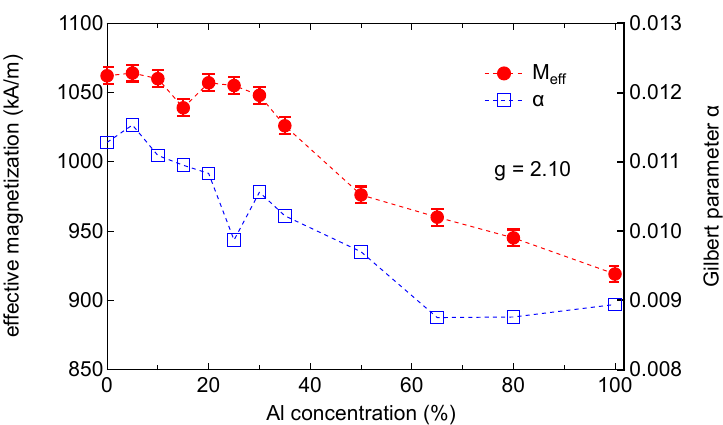}
\caption{\label{fig:FMR}
Effective magnetisation and Gilbert damping parameter $\alpha$ as a function of the Al concentration. The Lande-factor was held constant at $g=2.1$.}
\end{figure}

Broadband FMR measurements on unpatterned samples were performed up to 30\,GHz with the \textsc{OpenFMR} system. The Kittel analysis of the resonance positions and the Gilbert damping analysis of the resonance widths show substantial reductions with increasing Al content across the compositional series. The value of the effective magnetisation $M_\mathrm{eff}$ appears correlated with the crystallinity of the samples, with a plateau at low Al concentrations and a drop towards high Al concentration. This may point to either a reduction of the saturation magnetisation or an increase in the uniaxial out-of-plane anisotropy due to structural modifications of the films. The reduction of the Gilbert damping parameter $\alpha$ indicates a reduction of the spin mixing conductance as a result of reduced spin-orbit coupling and less effective spin absorption in the HEA with increasing Al concentration. Alternatively, this may also indicate a longer spin diffusion length with increasing Al concentration, which is not saturated at our sample thickness. A more detailed understanding would require film thickness series for every composition. For comparison, two reference samples show the following values: Ta 5.5nm / CFB 2.5nm, M$_\mathrm{eff} = 1064$\,kA/m and $\alpha = 0.013$; Pt 5.5nm / CFB 2.5nm, M$_\mathrm{eff} = 900$\,kA/m and $\alpha = 0.023$. Thus, our HEA series is, unsurprisingly, more similar to Ta films.

\begin{figure}[t!]
\includegraphics[width=8.6cm]{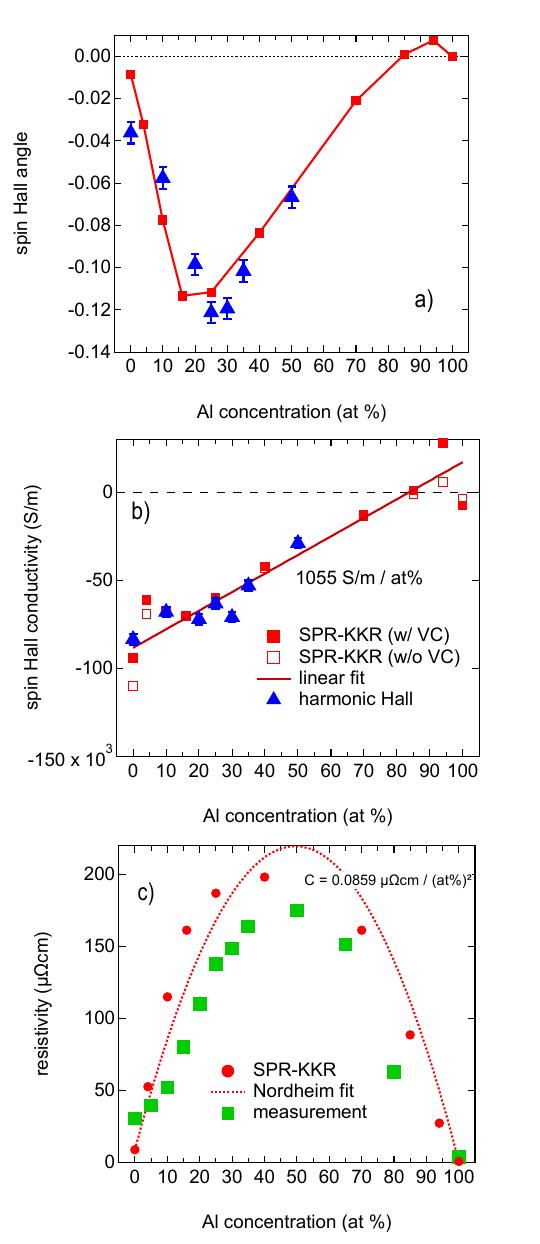}
\caption{\label{fig:electric}
Comparison of measured and calculated data. a) spin Hall conductivity, b) spin Hall angle, and c) resistivity. Red squares represent calculations, blue triangles are harmonic Hall results. In c), green filled squares represent four-point measurements. The Nordheim fit represents a parabolic fit with fixed endpoints to the SPR-KKR calculation results.}
\end{figure}

The harmonic Hall results for the SHA are compared to the ab initio-calculated SHA in Fig. \ref{fig:electric}\,a). The general trend agrees well and shows a peak of the SHA around the equiatomic composition. The SHC can be approximated by normalizing the observed torque in the harmonic Hall method directly to the electric field $E$ applied along the Hall cross, i.e. 
\begin{equation}
\sigma_\mathrm{SH} \approx \frac{2e}{\hbar} \frac{B_\mathrm{DL} M_\mathrm{s}t_\mathrm{CFB}}{E},
\end{equation}
where $B_\mathrm{DL}$ is obtained as a fit parameter from the harmonic Hall measurements and $M_\mathrm{s} \approx 1050$\,kA/m is the saturation magnetization of CoFeB. This method is easier to apply and more accurate than applying $\theta = \sigma_\mathrm{SH} \rho$ and applying a parallel resistor model. The SHC and the alloy resistivities are shown in Figure \ref{fig:electric}\,b) and c). Here, we observe an excellent matching of the harmonic Hall result and the ab initio-calculated SHC. Around the equiatomic composition, we find $\sigma_\mathrm{SH} \approx -72\,000\, \frac{\hbar}{2e}\,\mathrm{S/m}$, a spin Hall angle of $\theta \approx -0.12 \pm 0.01$, and a resistivity of $\rho \approx 140\,\mu\Omega$cm. These values are very similar to typical Ta thin films; however, here we obtain these for a crystalline bcc alloy, whereas Ta thin films are often amorphous. The SHA is also larger than that of our Pt reference sample, however the SHC is much larger in Pt. The SHC of the suspected Cr-(Mo,W) alloy is also close to the theoretical result of (Cr, Mo, W). This can be explained by looking at the calculated SHC of Cr and (Mo,W), which are $-30\,700\, \frac{\hbar}{2e}\,\mathrm{S/m}$ and $-112\,000\, \frac{\hbar}{2e}\,\mathrm{S/m}$, respectively. Then, the composition-weighted average is approximately $-84\,900\, \frac{\hbar}{2e}\,\mathrm{S/m}$, very similar to our experimental finding ($-83\,500\, \frac{\hbar}{2e}\,\mathrm{S/m}$) and slightly smaller than the calculated value of (Cr, Mo, W), $-94\,200\, \frac{\hbar}{2e}\,\mathrm{S/m}$. 

We emphasize that our calculations explicitly account for vertex corrections, i.e., extrinsic contributions (skew scattering and side-jump). As can be seen in Fig. \ref{fig:electric}\,b), these clearly play only a very small role in determining the total SHC of the alloy; in particular, for intermediate Al concentrations the results with and without vertex corrections are indistinguishable. Thus, it is a common misconception that the SHE in chemically disordered alloys would be dominated by extrinsic effects. Instead, it is solely given by the intrinsic contribution due to suppression of the vertex corrections in concentrated alloys.

Comparing the measured and ab initio-calculated resistivites, we observe that both follow the same trend: a huge enhancement of the resistivity towards 50\,\% Al concentration and the expected Nordheim-type inverted parabola. Our measurements, surprisingly, show slightly lower resistivity values compared to the calculated values. Usually, one would expect the opposite, due to finite-thickness effects, imperfect crystallization, grain boundary scattering, and static disorder contributions neglected in the calculation. Here, the slightly lower resistivity could indicate an inhomogeneous solid solution formation (i.e., a sub-regular solid solution \cite{Miracle2017}), where both locally Al-rich and Al-poor fractions would have lower resistivity, respectively. For the end-points of the composition series, we measured $\rho = 30.5\,\mu\Omega$cm for Cr-(Mo,W), and $\rho = 4.0\,\mu\Omega$cm for Al. The calculated value of $\rho = 8.9\,\mu\Omega$cm for CrMoW is much smaller than the measured value, but resistivity values around $30\,\mu\Omega$cm are quite typical for bcc transition metal thin films with our deposition conditions. The Nordheim fit is obtained as $\rho(x) = x\rho_0 + (1-x) \rho_{1} - C x(1-x)$. The magnitude of the parameter $C$ is outstanding: in typical alloys, only very modest enhancements of the resistivity can be observed. A resistivity of more than $150\,\mu\Omega$cm in a crystalline alloy is, thus, quite noteworthy. This can be explained along the lines of the resistivity theory in metals, which states that the impurity contribution to the resistivity can be roughly written as $\rho_\mathrm{imp}(x) \propto x(1-x) N(E_\mathrm{F}) (\Delta V)^2$. The density of states at the Fermi energy, $N(E_\mathrm{F}) \approx 0.85$\,states/eV remains mostly constant across the composition series and drops only towards the Al-rich side to $N(E_\mathrm{F}) \approx 0.55$\,states/eV. The scattering potential $\Delta V$ arises as the difference of the local scattering potentials with the surrounding matrix. In transition metal-main group alloys, this effect is naturally large due to the presence or lack of open $d$-shells, respectively. With maximum disorder, the resistivity is maximized, i.e. around the equiatomic composition. While trivial in binary alloys, in our quaternary alloy this leads to a skew of the calculated resistivity maximum towards lower Al concentration. However, the measurements reveal the maximum at 50\% Al concentration. This results has, however, to be taken with care due to the multiple phases present in the system, from low to high Al concentration: Cr + bcc-(Mo,W) $\rightarrow$ bcc-(Al, Cr, Mo, W) $\rightarrow$ amorphous $\rightarrow$ fcc-Al.

\section{Conclusion}
We have demonstrated the polycrystalline, textured growth of quarternary high-entropy alloy thin films. The choice of Ta as the seed layer is critical to obtain a homogenous bcc-crystallized phase around the equiatomic composition. A high resistivity of $140\,\mu\Omega$cm was achieved in the thin film of AlCrMoW. The spin Hall effect was characterized and a spin Hall conductivity of $-72\,000\, \frac{\hbar}{2e}\,\mathrm{S/m}$ and a spin Hall angle of $ -0.12 \pm 0.01$  were observed around the equiatomic composition. This work demonstrates the existence of high-resistivity high entropy alloys with a sizeable spin Hall effect. Further studies should investigate the possibility to enhance the phase homogeneity with thermal treatment and the epitaxial deposition of the alloy on single-crystalline substrates. Temperature dependent investigations will shed light on the electronic transport mechanism in the high-resistivity state. This will allow for further and deeper understanding of the spin Hall effect in this crystallograpically simple high resistivity alloy.

\section*{Appendix}
\subsection*{High Entropy Alloys}
High-entropy alloys (HEAs) are single-phase multi-element alloys stabilized by the configurational entropy.\cite{Miracle2017} The configurational entropy of mixing for an ideal solid solution is given by
\begin{equation}
\Delta S_\mathrm{mix} = -R \sum _{i=1}^N c_i \,\mathrm{ln}\left(c_i\right),
\end{equation}
where $R$ is the universal gas constant, $N$ is the number of constituents, and $c_i$ is the concentration of the $i$-th element. For equiatomic alloys, the configurational entropy is $\Delta S_\mathrm{mix} = R \, \mathrm{ln}(N)$. A value of $\Delta S_\mathrm{mix} > 1.5R$ (i.e. at least five components) is often associated with a high-entropy alloy, whereas broader definitions also include four-components alloys ($\Delta S_\mathrm{mix} = 1.39R$ for a equiatomic four-component alloy), emphasizing their stabilization by entropy (the \textit{high entropy effect} \cite{Miracle2017}). The stability of the HEA is assessed by comparing the Gibbs mixing energy $\Delta G_\mathrm{mix} = \Delta H_\mathrm{mix} - T\,\Delta S_\mathrm{mix}$ with the mixing enthalpy of the components, which is approximated as\cite{Zhang2008}
\begin{equation}
\Delta H_\mathrm{mix} = \sum_{i=1}^N \sum_{j>i}^N 4 \Delta H_{ij} c_i c_j,
\end{equation}
where $\Delta H_{ij}$ is the binary mixing enthalphy of elements $i$ and $j$. For the entropy effect to be dominant, the mixing enthalpy needs to fulfill $-15\,\mathrm{kJ/mol} < \Delta H_\mathrm{mix} < +5\,\mathrm{kJ/mol}$ (to avoid formation of binary phases or decomposition into the elements, respectively). The stability is usually expressed in terms of a thermodynamic parameter $\Omega$ as
\begin{equation}
\Omega = \frac{T_\mathrm{m} \Delta S_\mathrm{mix}}{\Delta H_\mathrm{mix}},
\end{equation}
where $T_\mathrm{m}$ is the average melting point of the constituent elements. For the formation of the HEA from the melt, $\Omega > 1.1$ is required. For the stability at lower temperature, additionally the Gibbs mixing energy at that temperature must remain lower than the formation energies of competing phases. Furthermore, the atomic sizes must be similar, quantified by a size mismatch parameter
\begin{equation}
\delta = \sqrt{ \sum _{i=1}^N c_i \left(1 - \frac{r_i}{\bar{r}}\right)^2 },
\end{equation}
where $R_i$ is the atomic radius of the $i$th atom and $\bar{r}$ is the concentration-weighted mean atomic radius. $\delta < 6.6\%$ is usually seen as favorable for the formation of a HEA. The valence electron concentration (VEC) per atom is indicative of the crystal structure of the solid solution, where VEC $< 6.87$ induces the bcc structure, VEC $> 8$ typically leads to the fcc structure, and intermediate values tend to result in phase mixtures. The required binary mixing enthalpies can be obtained from Ref. \cite{Takeuchi2005}.

\subsection*{X-Ray Reflectivity Fourier Transform}
We follow the method described in Ref. \cite{Sakurai2008}. We start with the raw x-ray reflectivity spectrum as a function of the scattering vector $q = 4\pi \sin \theta / \lambda$, with the angle $\theta = 2\theta / 2$ and the x-ray wavelength $\lambda$. The critical vector of the total reflection $q_\mathrm{c}$ is determined and a transformed axis $q' = \sqrt{q^2 - q_\mathrm{c}^2}$ is calculated. Using a log-log plot, a baseline of the type $I_\mathrm{b} = A/(q' - q'_0)^b$ is determined by a line fit. The oscillatory part is extracted as $I_\mathrm{osc} = I / I_\mathrm{b}$, and the mean value is subtracted to remove the zero-thickness component of the Fourier spectrum. Finally, a Hanning window and zero padding are applied and the real fast Fourier transform is applied to obtain the magnitude spectrum. The oscillatory part obtained with this procedure is harmonic and higher harmonics, which would give rise to false larger thicknesses, are not present. This was implemented in python with ChatGPT-5.

\begin{acknowledgments}
This work was supported by the Deutsche Forschungsgemeinschaft (DFG) under Project Numbers 513154775, 518575758, and under the Major Instrumentation Programme Project Numbers 468939474, and 511340083. We thank Prof. Alff (TU Darmstadt) for making available the x-ray diffractometer.

\end{acknowledgments}

\section*{AUTHOR DECLARATIONS }

\subsection*{Conflict of Interest}
The authors have no conflicts to disclose.

\section*{Data Availability Statement}

The data that support the findings of this study are available upon reasonable request from the corresponding author.

\section*{References}

\bibliography{cite} 

\begin{thebibliography}{34}%
\makeatletter
\providecommand \@ifxundefined [1]{%
 \@ifx{#1\undefined}
}%
\providecommand \@ifnum [1]{%
 \ifnum #1\expandafter \@firstoftwo
 \else \expandafter \@secondoftwo
 \fi
}%
\providecommand \@ifx [1]{%
 \ifx #1\expandafter \@firstoftwo
 \else \expandafter \@secondoftwo
 \fi
}%
\providecommand \natexlab [1]{#1}%
\providecommand \enquote  [1]{``#1''}%
\providecommand \bibnamefont  [1]{#1}%
\providecommand \bibfnamefont [1]{#1}%
\providecommand \citenamefont [1]{#1}%
\providecommand \href@noop [0]{\@secondoftwo}%
\providecommand \href [0]{\begingroup \@sanitize@url \@href}%
\providecommand \@href[1]{\@@startlink{#1}\@@href}%
\providecommand \@@href[1]{\endgroup#1\@@endlink}%
\providecommand \@sanitize@url [0]{\catcode `\\12\catcode `\$12\catcode
  `\&12\catcode `\#12\catcode `\^12\catcode `\_12\catcode `\%12\relax}%
\providecommand \@@startlink[1]{}%
\providecommand \@@endlink[0]{}%
\providecommand \url  [0]{\begingroup\@sanitize@url \@url }%
\providecommand \@url [1]{\endgroup\@href {#1}{\urlprefix }}%
\providecommand \urlprefix  [0]{URL }%
\providecommand \Eprint [0]{\href }%
\providecommand \doibase [0]{http://dx.doi.org/}%
\providecommand \selectlanguage [0]{\@gobble}%
\providecommand \bibinfo  [0]{\@secondoftwo}%
\providecommand \bibfield  [0]{\@secondoftwo}%
\providecommand \translation [1]{[#1]}%
\providecommand \BibitemOpen [0]{}%
\providecommand \bibitemStop [0]{}%
\providecommand \bibitemNoStop [0]{.\EOS\space}%
\providecommand \EOS [0]{\spacefactor3000\relax}%
\providecommand \BibitemShut  [1]{\csname bibitem#1\endcsname}%
\let\auto@bib@innerbib\@empty
\bibitem [{\citenamefont {Jungwirth}\ \emph {et~al.}(2012)\citenamefont
  {Jungwirth}, \citenamefont {Wunderlich},\ and\ \citenamefont
  {Olejník}}]{Jungwirth2012}%
  \BibitemOpen
  \bibfield  {author} {\bibinfo {author} {\bibfnamefont {T.}~\bibnamefont
  {Jungwirth}}, \bibinfo {author} {\bibfnamefont {J.}~\bibnamefont
  {Wunderlich}}, \ and\ \bibinfo {author} {\bibfnamefont {K.}~\bibnamefont
  {Olejník}},\ }\href {\doibase 10.1038/nmat3279} {\bibfield  {journal}
  {\bibinfo  {journal} {Nature Materials}\ }\textbf {\bibinfo {volume} {11}},\
  \bibinfo {pages} {382} (\bibinfo {year} {2012})}\BibitemShut {NoStop}%
\bibitem [{\citenamefont {Hoffmann}(2013)}]{Hoffmann2013}%
  \BibitemOpen
  \bibfield  {author} {\bibinfo {author} {\bibfnamefont {A.}~\bibnamefont
  {Hoffmann}},\ }\href {\doibase 10.1109/TMAG.2013.2262947} {\bibfield
  {journal} {\bibinfo  {journal} {IEEE Transactions on Magnetics}\ }\textbf
  {\bibinfo {volume} {49}},\ \bibinfo {pages} {5172} (\bibinfo {year}
  {2013})}\BibitemShut {NoStop}%
\bibitem [{\citenamefont {Ködderitzsch}\ \emph {et~al.}(2015)\citenamefont
  {Ködderitzsch}, \citenamefont {Chadova},\ and\ \citenamefont
  {Ebert}}]{Kdderitzsch2015}%
  \BibitemOpen
  \bibfield  {author} {\bibinfo {author} {\bibfnamefont {D.}~\bibnamefont
  {Ködderitzsch}}, \bibinfo {author} {\bibfnamefont {K.}~\bibnamefont
  {Chadova}}, \ and\ \bibinfo {author} {\bibfnamefont {H.}~\bibnamefont
  {Ebert}},\ }\href {\doibase 10.1103/PhysRevB.92.184415} {\bibfield  {journal}
  {\bibinfo  {journal} {Physical Review B}\ }\textbf {\bibinfo {volume} {92}},\
  \bibinfo {pages} {184415} (\bibinfo {year} {2015})}\BibitemShut {NoStop}%
\bibitem [{\citenamefont {Sagasta}\ \emph {et~al.}(2016)\citenamefont
  {Sagasta}, \citenamefont {Omori}, \citenamefont {Isasa}, \citenamefont
  {Gradhand}, \citenamefont {Hueso}, \citenamefont {Niimi}, \citenamefont
  {Otani},\ and\ \citenamefont {Casanova}}]{Sagasta2016}%
  \BibitemOpen
  \bibfield  {author} {\bibinfo {author} {\bibfnamefont {E.}~\bibnamefont
  {Sagasta}}, \bibinfo {author} {\bibfnamefont {Y.}~\bibnamefont {Omori}},
  \bibinfo {author} {\bibfnamefont {M.}~\bibnamefont {Isasa}}, \bibinfo
  {author} {\bibfnamefont {M.}~\bibnamefont {Gradhand}}, \bibinfo {author}
  {\bibfnamefont {L.~E.}\ \bibnamefont {Hueso}}, \bibinfo {author}
  {\bibfnamefont {Y.}~\bibnamefont {Niimi}}, \bibinfo {author} {\bibfnamefont
  {Y.}~\bibnamefont {Otani}}, \ and\ \bibinfo {author} {\bibfnamefont
  {F.}~\bibnamefont {Casanova}},\ }\href {\doibase 10.1103/PhysRevB.94.060412}
  {\bibfield  {journal} {\bibinfo  {journal} {Physical Review B}\ }\textbf
  {\bibinfo {volume} {94}},\ \bibinfo {pages} {060412} (\bibinfo {year}
  {2016})}\BibitemShut {NoStop}%
\bibitem [{\citenamefont {Qiao}\ \emph {et~al.}(2018)\citenamefont {Qiao},
  \citenamefont {Zhou}, \citenamefont {Yuan},\ and\ \citenamefont
  {Zhao}}]{Qiao2018}%
  \BibitemOpen
  \bibfield  {author} {\bibinfo {author} {\bibfnamefont {J.}~\bibnamefont
  {Qiao}}, \bibinfo {author} {\bibfnamefont {J.}~\bibnamefont {Zhou}}, \bibinfo
  {author} {\bibfnamefont {Z.}~\bibnamefont {Yuan}}, \ and\ \bibinfo {author}
  {\bibfnamefont {W.}~\bibnamefont {Zhao}},\ }\href {\doibase
  10.1103/PhysRevB.98.214402} {\bibfield  {journal} {\bibinfo  {journal}
  {Physical Review B}\ }\textbf {\bibinfo {volume} {98}},\ \bibinfo {pages}
  {214402} (\bibinfo {year} {2018})}\BibitemShut {NoStop}%
\bibitem [{\citenamefont {McHugh}\ \emph {et~al.}(2020)\citenamefont {McHugh},
  \citenamefont {Goh}, \citenamefont {Gradhand},\ and\ \citenamefont
  {Stewart}}]{McHugh2020}%
  \BibitemOpen
  \bibfield  {author} {\bibinfo {author} {\bibfnamefont {O.~L.}\ \bibnamefont
  {McHugh}}, \bibinfo {author} {\bibfnamefont {W.~F.}\ \bibnamefont {Goh}},
  \bibinfo {author} {\bibfnamefont {M.}~\bibnamefont {Gradhand}}, \ and\
  \bibinfo {author} {\bibfnamefont {D.~A.}\ \bibnamefont {Stewart}},\ }\href
  {\doibase 10.1103/PhysRevMaterials.4.094404} {\bibfield  {journal} {\bibinfo
  {journal} {Physical Review Materials}\ }\textbf {\bibinfo {volume} {4}}
  (\bibinfo {year} {2020}),\ 10.1103/PhysRevMaterials.4.094404}\BibitemShut
  {NoStop}%
\bibitem [{\citenamefont {Nguyen}\ \emph {et~al.}(2016)\citenamefont {Nguyen},
  \citenamefont {Ralph},\ and\ \citenamefont {Buhrman}}]{Nguyen2016}%
  \BibitemOpen
  \bibfield  {author} {\bibinfo {author} {\bibfnamefont {M.-H.}\ \bibnamefont
  {Nguyen}}, \bibinfo {author} {\bibfnamefont {D.~C.}\ \bibnamefont {Ralph}}, \
  and\ \bibinfo {author} {\bibfnamefont {R.~A.}\ \bibnamefont {Buhrman}},\
  }\href {\doibase 10.1103/PhysRevLett.116.126601} {\bibfield  {journal}
  {\bibinfo  {journal} {Physical Review Letters}\ }\textbf {\bibinfo {volume}
  {116}},\ \bibinfo {pages} {126601} (\bibinfo {year} {2016})}\BibitemShut
  {NoStop}%
\bibitem [{\citenamefont {Dutta}\ \emph {et~al.}(2017)\citenamefont {Dutta},
  \citenamefont {Sankaran}, \citenamefont {Moors}, \citenamefont {Pourtois},
  \citenamefont {Elshocht}, \citenamefont {Bömmels}, \citenamefont
  {Vandervorst}, \citenamefont {Tőkei},\ and\ \citenamefont
  {Adelmann}}]{Dutta2017}%
  \BibitemOpen
  \bibfield  {author} {\bibinfo {author} {\bibfnamefont {S.}~\bibnamefont
  {Dutta}}, \bibinfo {author} {\bibfnamefont {K.}~\bibnamefont {Sankaran}},
  \bibinfo {author} {\bibfnamefont {K.}~\bibnamefont {Moors}}, \bibinfo
  {author} {\bibfnamefont {G.}~\bibnamefont {Pourtois}}, \bibinfo {author}
  {\bibfnamefont {S.~V.}\ \bibnamefont {Elshocht}}, \bibinfo {author}
  {\bibfnamefont {J.}~\bibnamefont {Bömmels}}, \bibinfo {author}
  {\bibfnamefont {W.}~\bibnamefont {Vandervorst}}, \bibinfo {author}
  {\bibfnamefont {Z.}~\bibnamefont {Tőkei}}, \ and\ \bibinfo {author}
  {\bibfnamefont {C.}~\bibnamefont {Adelmann}},\ }\href {\doibase
  10.1063/1.4992089} {\bibfield  {journal} {\bibinfo  {journal} {Journal of
  Applied Physics}\ }\textbf {\bibinfo {volume} {122}},\ \bibinfo {pages}
  {025107} (\bibinfo {year} {2017})}\BibitemShut {NoStop}%
\bibitem [{\citenamefont {Hao}\ \emph {et~al.}(2015)\citenamefont {Hao},
  \citenamefont {Chen},\ and\ \citenamefont {Xiao}}]{Hao2015}%
  \BibitemOpen
  \bibfield  {author} {\bibinfo {author} {\bibfnamefont {Q.}~\bibnamefont
  {Hao}}, \bibinfo {author} {\bibfnamefont {W.}~\bibnamefont {Chen}}, \ and\
  \bibinfo {author} {\bibfnamefont {G.}~\bibnamefont {Xiao}},\ }\href {\doibase
  10.1063/1.4919867} {\bibfield  {journal} {\bibinfo  {journal} {Applied
  Physics Letters}\ }\textbf {\bibinfo {volume} {106}},\ \bibinfo {pages}
  {182403} (\bibinfo {year} {2015})}\BibitemShut {NoStop}%
\bibitem [{\citenamefont {Kim}\ \emph {et~al.}(2020)\citenamefont {Kim},
  \citenamefont {Han}, \citenamefont {Vafaee}, \citenamefont {Jaiswal},
  \citenamefont {Lee}, \citenamefont {Jakob},\ and\ \citenamefont
  {Kläui}}]{Kim2020}%
  \BibitemOpen
  \bibfield  {author} {\bibinfo {author} {\bibfnamefont {J.-Y.}\ \bibnamefont
  {Kim}}, \bibinfo {author} {\bibfnamefont {D.-S.}\ \bibnamefont {Han}},
  \bibinfo {author} {\bibfnamefont {M.}~\bibnamefont {Vafaee}}, \bibinfo
  {author} {\bibfnamefont {S.}~\bibnamefont {Jaiswal}}, \bibinfo {author}
  {\bibfnamefont {K.}~\bibnamefont {Lee}}, \bibinfo {author} {\bibfnamefont
  {G.}~\bibnamefont {Jakob}}, \ and\ \bibinfo {author} {\bibfnamefont
  {M.}~\bibnamefont {Kläui}},\ }\href {\doibase 10.1063/5.0022012} {\bibfield
  {journal} {\bibinfo  {journal} {Applied Physics Letters}\ }\textbf {\bibinfo
  {volume} {117}} (\bibinfo {year} {2020}),\ 10.1063/5.0022012}\BibitemShut
  {NoStop}%
\bibitem [{\citenamefont {Demasius}\ \emph {et~al.}(2016)\citenamefont
  {Demasius}, \citenamefont {Phung}, \citenamefont {Zhang}, \citenamefont
  {Hughes}, \citenamefont {Yang}, \citenamefont {Kellock}, \citenamefont {Han},
  \citenamefont {Pushp},\ and\ \citenamefont {Parkin}}]{Demasius2016}%
  \BibitemOpen
  \bibfield  {author} {\bibinfo {author} {\bibfnamefont {K.-U.}\ \bibnamefont
  {Demasius}}, \bibinfo {author} {\bibfnamefont {T.}~\bibnamefont {Phung}},
  \bibinfo {author} {\bibfnamefont {W.}~\bibnamefont {Zhang}}, \bibinfo
  {author} {\bibfnamefont {B.~P.}\ \bibnamefont {Hughes}}, \bibinfo {author}
  {\bibfnamefont {S.-H.}\ \bibnamefont {Yang}}, \bibinfo {author}
  {\bibfnamefont {A.}~\bibnamefont {Kellock}}, \bibinfo {author} {\bibfnamefont
  {W.}~\bibnamefont {Han}}, \bibinfo {author} {\bibfnamefont {A.}~\bibnamefont
  {Pushp}}, \ and\ \bibinfo {author} {\bibfnamefont {S.~S.~P.}\ \bibnamefont
  {Parkin}},\ }\href {\doibase 10.1038/ncomms10644} {\bibfield  {journal}
  {\bibinfo  {journal} {Nature Communications}\ }\textbf {\bibinfo {volume}
  {7}},\ \bibinfo {pages} {10644} (\bibinfo {year} {2016})}\BibitemShut
  {NoStop}%
\bibitem [{\citenamefont {Neumann}\ \emph {et~al.}(2016)\citenamefont
  {Neumann}, \citenamefont {Meier}, \citenamefont {Schmalhorst}, \citenamefont
  {Rott}, \citenamefont {Reiss},\ and\ \citenamefont {Meinert}}]{Neumann2016}%
  \BibitemOpen
  \bibfield  {author} {\bibinfo {author} {\bibfnamefont {L.}~\bibnamefont
  {Neumann}}, \bibinfo {author} {\bibfnamefont {D.}~\bibnamefont {Meier}},
  \bibinfo {author} {\bibfnamefont {J.}~\bibnamefont {Schmalhorst}}, \bibinfo
  {author} {\bibfnamefont {K.}~\bibnamefont {Rott}}, \bibinfo {author}
  {\bibfnamefont {G.}~\bibnamefont {Reiss}}, \ and\ \bibinfo {author}
  {\bibfnamefont {M.}~\bibnamefont {Meinert}},\ }\href {\doibase
  10.1063/1.4964415} {\bibfield  {journal} {\bibinfo  {journal} {Applied
  Physics Letters}\ }\textbf {\bibinfo {volume} {109}},\ \bibinfo {pages}
  {142405} (\bibinfo {year} {2016})}\BibitemShut {NoStop}%
\bibitem [{\citenamefont {Sethu}\ \emph {et~al.}(2021)\citenamefont {Sethu},
  \citenamefont {Ghosh}, \citenamefont {Couet}, \citenamefont {Swerts},
  \citenamefont {Sorée}, \citenamefont {Boeck}, \citenamefont {Kar},\ and\
  \citenamefont {Garello}}]{Sethu2021}%
  \BibitemOpen
  \bibfield  {author} {\bibinfo {author} {\bibfnamefont {K.~K.~V.}\
  \bibnamefont {Sethu}}, \bibinfo {author} {\bibfnamefont {S.}~\bibnamefont
  {Ghosh}}, \bibinfo {author} {\bibfnamefont {S.}~\bibnamefont {Couet}},
  \bibinfo {author} {\bibfnamefont {J.}~\bibnamefont {Swerts}}, \bibinfo
  {author} {\bibfnamefont {B.}~\bibnamefont {Sorée}}, \bibinfo {author}
  {\bibfnamefont {J.~D.}\ \bibnamefont {Boeck}}, \bibinfo {author}
  {\bibfnamefont {G.~S.}\ \bibnamefont {Kar}}, \ and\ \bibinfo {author}
  {\bibfnamefont {K.}~\bibnamefont {Garello}},\ }\href {\doibase
  10.1103/PhysRevApplied.16.064009} {\bibfield  {journal} {\bibinfo  {journal}
  {Physical Review Applied}\ }\textbf {\bibinfo {volume} {16}},\ \bibinfo
  {pages} {064009} (\bibinfo {year} {2021})}\BibitemShut {NoStop}%
\bibitem [{\citenamefont {Gómez}\ and\ \citenamefont
  {Haberkorn}(2023)}]{Gmez2023}%
  \BibitemOpen
  \bibfield  {author} {\bibinfo {author} {\bibfnamefont {J.~E.}\ \bibnamefont
  {Gómez}}\ and\ \bibinfo {author} {\bibfnamefont {N.}~\bibnamefont
  {Haberkorn}},\ }\href {\doibase 10.1007/s00339-023-06695-x} {\bibfield
  {journal} {\bibinfo  {journal} {Applied Physics A: Materials Science and
  Processing}\ }\textbf {\bibinfo {volume} {129}} (\bibinfo {year} {2023}),\
  10.1007/s00339-023-06695-x}\BibitemShut {NoStop}%
\bibitem [{\citenamefont {Sui}\ \emph {et~al.}(2017)\citenamefont {Sui},
  \citenamefont {Wang}, \citenamefont {Kim}, \citenamefont {Wang},
  \citenamefont {Rhim}, \citenamefont {Duan},\ and\ \citenamefont
  {Kioussis}}]{Sui2017}%
  \BibitemOpen
  \bibfield  {author} {\bibinfo {author} {\bibfnamefont {X.}~\bibnamefont
  {Sui}}, \bibinfo {author} {\bibfnamefont {C.}~\bibnamefont {Wang}}, \bibinfo
  {author} {\bibfnamefont {J.}~\bibnamefont {Kim}}, \bibinfo {author}
  {\bibfnamefont {J.}~\bibnamefont {Wang}}, \bibinfo {author} {\bibfnamefont
  {S.~H.}\ \bibnamefont {Rhim}}, \bibinfo {author} {\bibfnamefont
  {W.}~\bibnamefont {Duan}}, \ and\ \bibinfo {author} {\bibfnamefont
  {N.}~\bibnamefont {Kioussis}},\ }\href {\doibase 10.1103/PhysRevB.96.241105}
  {\bibfield  {journal} {\bibinfo  {journal} {Physical Review B}\ }\textbf
  {\bibinfo {volume} {96}},\ \bibinfo {pages} {241105} (\bibinfo {year}
  {2017})}\BibitemShut {NoStop}%
\bibitem [{\citenamefont {Janus}\ \emph
  {et~al.}(2025{\natexlab{a}})\citenamefont {Janus}, \citenamefont {Yadav},
  \citenamefont {Beermann}, \citenamefont {Zhang}, \citenamefont {Hafez},
  \citenamefont {Turchinovich}, \citenamefont {Preu},\ and\ \citenamefont
  {Meinert}}]{Janus2025a}%
  \BibitemOpen
  \bibfield  {author} {\bibinfo {author} {\bibfnamefont {F.}~\bibnamefont
  {Janus}}, \bibinfo {author} {\bibfnamefont {J.}~\bibnamefont {Yadav}},
  \bibinfo {author} {\bibfnamefont {N.~S.}\ \bibnamefont {Beermann}}, \bibinfo
  {author} {\bibfnamefont {W.}~\bibnamefont {Zhang}}, \bibinfo {author}
  {\bibfnamefont {H.~A.}\ \bibnamefont {Hafez}}, \bibinfo {author}
  {\bibfnamefont {D.}~\bibnamefont {Turchinovich}}, \bibinfo {author}
  {\bibfnamefont {S.}~\bibnamefont {Preu}}, \ and\ \bibinfo {author}
  {\bibfnamefont {M.}~\bibnamefont {Meinert}},\ }\href {\doibase
  10.1103/PhysRevB.111.174411} {\bibfield  {journal} {\bibinfo  {journal}
  {Physical Review B}\ }\textbf {\bibinfo {volume} {111}},\ \bibinfo {pages}
  {174411} (\bibinfo {year} {2025}{\natexlab{a}})}\BibitemShut {NoStop}%
\bibitem [{\citenamefont {Obstbaum}\ \emph {et~al.}(2016)\citenamefont
  {Obstbaum}, \citenamefont {Decker}, \citenamefont {Greitner}, \citenamefont
  {Haertinger}, \citenamefont {Meier}, \citenamefont {Kronseder}, \citenamefont
  {Chadova}, \citenamefont {Wimmer}, \citenamefont {Ködderitzsch},
  \citenamefont {Ebert},\ and\ \citenamefont {Back}}]{Obstbaum2016}%
  \BibitemOpen
  \bibfield  {author} {\bibinfo {author} {\bibfnamefont {M.}~\bibnamefont
  {Obstbaum}}, \bibinfo {author} {\bibfnamefont {M.}~\bibnamefont {Decker}},
  \bibinfo {author} {\bibfnamefont {A.~K.}\ \bibnamefont {Greitner}}, \bibinfo
  {author} {\bibfnamefont {M.}~\bibnamefont {Haertinger}}, \bibinfo {author}
  {\bibfnamefont {T.~N.~G.}\ \bibnamefont {Meier}}, \bibinfo {author}
  {\bibfnamefont {M.}~\bibnamefont {Kronseder}}, \bibinfo {author}
  {\bibfnamefont {K.}~\bibnamefont {Chadova}}, \bibinfo {author} {\bibfnamefont
  {S.}~\bibnamefont {Wimmer}}, \bibinfo {author} {\bibfnamefont
  {D.}~\bibnamefont {Ködderitzsch}}, \bibinfo {author} {\bibfnamefont
  {H.}~\bibnamefont {Ebert}}, \ and\ \bibinfo {author} {\bibfnamefont {C.~H.}\
  \bibnamefont {Back}},\ }\href {\doibase 10.1103/PhysRevLett.117.167204}
  {\bibfield  {journal} {\bibinfo  {journal} {Physical Review Letters}\
  }\textbf {\bibinfo {volume} {117}},\ \bibinfo {pages} {167204} (\bibinfo
  {year} {2016})}\BibitemShut {NoStop}%
\bibitem [{\citenamefont {Meinert}\ \emph {et~al.}(2020)\citenamefont
  {Meinert}, \citenamefont {Gliniors}, \citenamefont {Gueckstock},
  \citenamefont {Seifert}, \citenamefont {Liensberger}, \citenamefont {Weiler},
  \citenamefont {Wimmer}, \citenamefont {Ebert},\ and\ \citenamefont
  {Kampfrath}}]{Meinert2020}%
  \BibitemOpen
  \bibfield  {author} {\bibinfo {author} {\bibfnamefont {M.}~\bibnamefont
  {Meinert}}, \bibinfo {author} {\bibfnamefont {B.}~\bibnamefont {Gliniors}},
  \bibinfo {author} {\bibfnamefont {O.}~\bibnamefont {Gueckstock}}, \bibinfo
  {author} {\bibfnamefont {T.~S.}\ \bibnamefont {Seifert}}, \bibinfo {author}
  {\bibfnamefont {L.}~\bibnamefont {Liensberger}}, \bibinfo {author}
  {\bibfnamefont {M.}~\bibnamefont {Weiler}}, \bibinfo {author} {\bibfnamefont
  {S.}~\bibnamefont {Wimmer}}, \bibinfo {author} {\bibfnamefont
  {H.}~\bibnamefont {Ebert}}, \ and\ \bibinfo {author} {\bibfnamefont
  {T.}~\bibnamefont {Kampfrath}},\ }\href {\doibase
  10.1103/PhysRevApplied.14.064011} {\bibfield  {journal} {\bibinfo  {journal}
  {Physical Review Applied}\ }\textbf {\bibinfo {volume} {14}},\ \bibinfo
  {pages} {064011} (\bibinfo {year} {2020})}\BibitemShut {NoStop}%
\bibitem [{\citenamefont {McHugh}\ \emph {et~al.}(2024)\citenamefont {McHugh},
  \citenamefont {Gradhand},\ and\ \citenamefont {Stewart}}]{McHugh2024}%
  \BibitemOpen
  \bibfield  {author} {\bibinfo {author} {\bibfnamefont {O.~L.}\ \bibnamefont
  {McHugh}}, \bibinfo {author} {\bibfnamefont {M.}~\bibnamefont {Gradhand}}, \
  and\ \bibinfo {author} {\bibfnamefont {D.~A.}\ \bibnamefont {Stewart}},\
  }\href {\doibase 10.1103/PhysRevMaterials.8.015003} {\bibfield  {journal}
  {\bibinfo  {journal} {Physical Review Materials}\ }\textbf {\bibinfo {volume}
  {8}} (\bibinfo {year} {2024}),\
  10.1103/PhysRevMaterials.8.015003}\BibitemShut {NoStop}%
\bibitem [{\citenamefont {Janus}\ \emph
  {et~al.}(2025{\natexlab{b}})\citenamefont {Janus}, \citenamefont {Beermann},
  \citenamefont {Yadav}, \citenamefont {Lekha}, \citenamefont {Zhang},
  \citenamefont {Hafez}, \citenamefont {Turchinovich},\ and\ \citenamefont
  {Meinert}}]{Janus2025b}%
  \BibitemOpen
  \bibfield  {author} {\bibinfo {author} {\bibfnamefont {F.}~\bibnamefont
  {Janus}}, \bibinfo {author} {\bibfnamefont {N.}~\bibnamefont {Beermann}},
  \bibinfo {author} {\bibfnamefont {J.}~\bibnamefont {Yadav}}, \bibinfo
  {author} {\bibfnamefont {R.~R.}\ \bibnamefont {Lekha}}, \bibinfo {author}
  {\bibfnamefont {W.}~\bibnamefont {Zhang}}, \bibinfo {author} {\bibfnamefont
  {H.~A.}\ \bibnamefont {Hafez}}, \bibinfo {author} {\bibfnamefont
  {D.}~\bibnamefont {Turchinovich}}, \ and\ \bibinfo {author} {\bibfnamefont
  {M.}~\bibnamefont {Meinert}},\ }\href {https://arxiv.org/abs/2504.07614}
  {\enquote {\bibinfo {title} {Enhanced thz emission from spintronic emitters
  with pt-al alloys},}\ } (\bibinfo {year} {2025}{\natexlab{b}}),\ \Eprint
  {http://arxiv.org/abs/2504.07614} {arXiv:2504.07614 [cond-mat.mtrl-sci]}
  \BibitemShut {NoStop}%
\bibitem [{\citenamefont {Miracle}\ and\ \citenamefont
  {Senkov}(2017)}]{Miracle2017}%
  \BibitemOpen
  \bibfield  {author} {\bibinfo {author} {\bibfnamefont {D.}~\bibnamefont
  {Miracle}}\ and\ \bibinfo {author} {\bibfnamefont {O.}~\bibnamefont
  {Senkov}},\ }\href {\doibase 10.1016/j.actamat.2016.08.081} {\bibfield
  {journal} {\bibinfo  {journal} {Acta Materialia}\ }\textbf {\bibinfo {volume}
  {122}},\ \bibinfo {pages} {448} (\bibinfo {year} {2017})}\BibitemShut
  {NoStop}%
\bibitem [{\citenamefont {Rittiruam}\ \emph {et~al.}(2022)\citenamefont
  {Rittiruam}, \citenamefont {Noppakhun}, \citenamefont {Setasuban},
  \citenamefont {Aumnongpho}, \citenamefont {Sriwattana}, \citenamefont
  {Boonchuay}, \citenamefont {Saelee}, \citenamefont {Wangphon}, \citenamefont
  {Ektarawong}, \citenamefont {Chammingkwan}, \citenamefont {Taniike},
  \citenamefont {Praserthdam},\ and\ \citenamefont
  {Praserthdam}}]{Rittiruam2022}%
  \BibitemOpen
  \bibfield  {author} {\bibinfo {author} {\bibfnamefont {M.}~\bibnamefont
  {Rittiruam}}, \bibinfo {author} {\bibfnamefont {J.}~\bibnamefont
  {Noppakhun}}, \bibinfo {author} {\bibfnamefont {S.}~\bibnamefont
  {Setasuban}}, \bibinfo {author} {\bibfnamefont {N.}~\bibnamefont
  {Aumnongpho}}, \bibinfo {author} {\bibfnamefont {A.}~\bibnamefont
  {Sriwattana}}, \bibinfo {author} {\bibfnamefont {S.}~\bibnamefont
  {Boonchuay}}, \bibinfo {author} {\bibfnamefont {T.}~\bibnamefont {Saelee}},
  \bibinfo {author} {\bibfnamefont {C.}~\bibnamefont {Wangphon}}, \bibinfo
  {author} {\bibfnamefont {A.}~\bibnamefont {Ektarawong}}, \bibinfo {author}
  {\bibfnamefont {P.}~\bibnamefont {Chammingkwan}}, \bibinfo {author}
  {\bibfnamefont {T.}~\bibnamefont {Taniike}}, \bibinfo {author} {\bibfnamefont
  {S.}~\bibnamefont {Praserthdam}}, \ and\ \bibinfo {author} {\bibfnamefont
  {P.}~\bibnamefont {Praserthdam}},\ }\href {\doibase
  10.1038/s41598-022-21209-0} {\bibfield  {journal} {\bibinfo  {journal}
  {Scientific Reports}\ }\textbf {\bibinfo {volume} {12}},\ \bibinfo {pages}
  {16653} (\bibinfo {year} {2022})}\BibitemShut {NoStop}%
\bibitem [{\citenamefont {Chen}\ \emph {et~al.}(2017)\citenamefont {Chen},
  \citenamefont {Chuang}, \citenamefont {Huang}, \citenamefont {Yen},\ and\
  \citenamefont {Pai}}]{Chen2017}%
  \BibitemOpen
  \bibfield  {author} {\bibinfo {author} {\bibfnamefont {T.~Y.}\ \bibnamefont
  {Chen}}, \bibinfo {author} {\bibfnamefont {T.~C.}\ \bibnamefont {Chuang}},
  \bibinfo {author} {\bibfnamefont {S.~Y.}\ \bibnamefont {Huang}}, \bibinfo
  {author} {\bibfnamefont {H.~W.}\ \bibnamefont {Yen}}, \ and\ \bibinfo
  {author} {\bibfnamefont {C.~F.}\ \bibnamefont {Pai}},\ }\href {\doibase
  10.1103/PhysRevApplied.8.044005} {\bibfield  {journal} {\bibinfo  {journal}
  {Physical Review Applied}\ }\textbf {\bibinfo {volume} {8}} (\bibinfo {year}
  {2017}),\ 10.1103/PhysRevApplied.8.044005}\BibitemShut {NoStop}%
\bibitem [{\citenamefont {Kubota}\ \emph {et~al.}(2025)\citenamefont {Kubota},
  \citenamefont {Suzuki}, \citenamefont {Hirayama}, \citenamefont {Takahashi},\
  and\ \citenamefont {Takanashi}}]{Kubota2025}%
  \BibitemOpen
  \bibfield  {author} {\bibinfo {author} {\bibfnamefont {T.}~\bibnamefont
  {Kubota}}, \bibinfo {author} {\bibfnamefont {K.~Z.}\ \bibnamefont {Suzuki}},
  \bibinfo {author} {\bibfnamefont {Y.}~\bibnamefont {Hirayama}}, \bibinfo
  {author} {\bibfnamefont {S.}~\bibnamefont {Takahashi}}, \ and\ \bibinfo
  {author} {\bibfnamefont {K.}~\bibnamefont {Takanashi}},\ }\href
  {http://arxiv.org/abs/2503.03218} {\  (\bibinfo {year} {2025})}\BibitemShut
  {NoStop}%
\bibitem [{\citenamefont {Wang}\ \emph {et~al.}(2025)\citenamefont {Wang},
  \citenamefont {Migliorini}, \citenamefont {Li}, \citenamefont {Deniz},
  \citenamefont {Kostanovski}, \citenamefont {Jeon},\ and\ \citenamefont
  {Parkin}}]{Wang2025}%
  \BibitemOpen
  \bibfield  {author} {\bibinfo {author} {\bibfnamefont {P.}~\bibnamefont
  {Wang}}, \bibinfo {author} {\bibfnamefont {A.}~\bibnamefont {Migliorini}},
  \bibinfo {author} {\bibfnamefont {Y.}~\bibnamefont {Li}}, \bibinfo {author}
  {\bibfnamefont {H.}~\bibnamefont {Deniz}}, \bibinfo {author} {\bibfnamefont
  {I.}~\bibnamefont {Kostanovski}}, \bibinfo {author} {\bibfnamefont
  {J.}~\bibnamefont {Jeon}}, \ and\ \bibinfo {author} {\bibfnamefont
  {S.~S.~P.}\ \bibnamefont {Parkin}},\ }\href {\doibase 10.1002/adma.202416820}
  {\bibfield  {journal} {\bibinfo  {journal} {Advanced Materials}\ }\textbf
  {\bibinfo {volume} {37}} (\bibinfo {year} {2025}),\
  10.1002/adma.202416820}\BibitemShut {NoStop}%
\bibitem [{\citenamefont {Fukushima}\ \emph {et~al.}(2022)\citenamefont
  {Fukushima}, \citenamefont {Akai}, \citenamefont {Chikyow},\ and\
  \citenamefont {Kino}}]{Fukushima2022}%
  \BibitemOpen
  \bibfield  {author} {\bibinfo {author} {\bibfnamefont {T.}~\bibnamefont
  {Fukushima}}, \bibinfo {author} {\bibfnamefont {H.}~\bibnamefont {Akai}},
  \bibinfo {author} {\bibfnamefont {T.}~\bibnamefont {Chikyow}}, \ and\
  \bibinfo {author} {\bibfnamefont {H.}~\bibnamefont {Kino}},\ }\href {\doibase
  10.1103/PhysRevMaterials.6.023802} {\bibfield  {journal} {\bibinfo  {journal}
  {Physical Review Materials}\ }\textbf {\bibinfo {volume} {6}},\ \bibinfo
  {pages} {023802} (\bibinfo {year} {2022})}\BibitemShut {NoStop}%
\bibitem [{\citenamefont {Go}\ \emph {et~al.}(2024)\citenamefont {Go},
  \citenamefont {Lee}, \citenamefont {Oppeneer}, \citenamefont {Blügel},\ and\
  \citenamefont {Mokrousov}}]{Go2024}%
  \BibitemOpen
  \bibfield  {author} {\bibinfo {author} {\bibfnamefont {D.}~\bibnamefont
  {Go}}, \bibinfo {author} {\bibfnamefont {H.-W.}\ \bibnamefont {Lee}},
  \bibinfo {author} {\bibfnamefont {P.~M.}\ \bibnamefont {Oppeneer}}, \bibinfo
  {author} {\bibfnamefont {S.}~\bibnamefont {Blügel}}, \ and\ \bibinfo
  {author} {\bibfnamefont {Y.}~\bibnamefont {Mokrousov}},\ }\href {\doibase
  10.1103/PhysRevB.109.174435} {\bibfield  {journal} {\bibinfo  {journal}
  {Physical Review B}\ }\textbf {\bibinfo {volume} {109}},\ \bibinfo {pages}
  {174435} (\bibinfo {year} {2024})}\BibitemShut {NoStop}%
\bibitem [{\citenamefont {de~Oliveira~Schneider}\ \emph
  {et~al.}(2025)\citenamefont {de~Oliveira~Schneider}, \citenamefont {Sharma},
  \citenamefont {Khan},\ and\ \citenamefont
  {Meinert}}]{deOliveiraSchneider2025}%
  \BibitemOpen
  \bibfield  {author} {\bibinfo {author} {\bibfnamefont {T.}~\bibnamefont
  {de~Oliveira~Schneider}}, \bibinfo {author} {\bibfnamefont {S.}~\bibnamefont
  {Sharma}}, \bibinfo {author} {\bibfnamefont {A.}~\bibnamefont {Khan}}, \ and\
  \bibinfo {author} {\bibfnamefont {M.}~\bibnamefont {Meinert}},\ }\href
  {\doibase 10.1063/5.0241406} {\bibfield  {journal} {\bibinfo  {journal}
  {Review of Scientific Instruments}\ }\textbf {\bibinfo {volume} {96}}
  (\bibinfo {year} {2025}),\ 10.1063/5.0241406}\BibitemShut {NoStop}%
\bibitem [{\citenamefont {Liu}\ \emph {et~al.}(2011)\citenamefont {Liu},
  \citenamefont {Moriyama}, \citenamefont {Ralph},\ and\ \citenamefont
  {Buhrman}}]{Liu2011}%
  \BibitemOpen
  \bibfield  {author} {\bibinfo {author} {\bibfnamefont {L.}~\bibnamefont
  {Liu}}, \bibinfo {author} {\bibfnamefont {T.}~\bibnamefont {Moriyama}},
  \bibinfo {author} {\bibfnamefont {D.~C.}\ \bibnamefont {Ralph}}, \ and\
  \bibinfo {author} {\bibfnamefont {R.~A.}\ \bibnamefont {Buhrman}},\ }\href
  {\doibase 10.1103/PhysRevLett.106.036601} {\bibfield  {journal} {\bibinfo
  {journal} {Physical Review Letters}\ }\textbf {\bibinfo {volume} {106}},\
  \bibinfo {pages} {036601} (\bibinfo {year} {2011})}\BibitemShut {NoStop}%
\bibitem [{\citenamefont {Fritz}\ \emph {et~al.}(2018)\citenamefont {Fritz},
  \citenamefont {Wimmer}, \citenamefont {Ebert},\ and\ \citenamefont
  {Meinert}}]{Fritz2018}%
  \BibitemOpen
  \bibfield  {author} {\bibinfo {author} {\bibfnamefont {K.}~\bibnamefont
  {Fritz}}, \bibinfo {author} {\bibfnamefont {S.}~\bibnamefont {Wimmer}},
  \bibinfo {author} {\bibfnamefont {H.}~\bibnamefont {Ebert}}, \ and\ \bibinfo
  {author} {\bibfnamefont {M.}~\bibnamefont {Meinert}},\ }\href {\doibase
  10.1103/PhysRevB.98.094433} {\bibfield  {journal} {\bibinfo  {journal}
  {Physical Review B}\ }\textbf {\bibinfo {volume} {98}},\ \bibinfo {pages}
  {094433} (\bibinfo {year} {2018})}\BibitemShut {NoStop}%
\bibitem [{\citenamefont {Sakurai}\ \emph {et~al.}(2008)\citenamefont
  {Sakurai}, \citenamefont {Mizusawa},\ and\ \citenamefont
  {Ishii}}]{Sakurai2008}%
  \BibitemOpen
  \bibfield  {author} {\bibinfo {author} {\bibfnamefont {K.}~\bibnamefont
  {Sakurai}}, \bibinfo {author} {\bibfnamefont {M.}~\bibnamefont {Mizusawa}}, \
  and\ \bibinfo {author} {\bibfnamefont {M.}~\bibnamefont {Ishii}},\ }\href
  {\doibase 10.14723/tmrsj.33.523} {\bibfield  {journal} {\bibinfo  {journal}
  {Transactions of the Materials Research Society of Japan}\ }\textbf {\bibinfo
  {volume} {33}},\ \bibinfo {pages} {523} (\bibinfo {year} {2008})}\BibitemShut
  {NoStop}%
\bibitem [{\citenamefont {Liu}\ \emph {et~al.}(2021)\citenamefont {Liu},
  \citenamefont {Zhang}, \citenamefont {Sun}, \citenamefont {Miao},
  \citenamefont {Wang},\ and\ \citenamefont {Ding}}]{Liu2021}%
  \BibitemOpen
  \bibfield  {author} {\bibinfo {author} {\bibfnamefont {Q.}~\bibnamefont
  {Liu}}, \bibinfo {author} {\bibfnamefont {Y.}~\bibnamefont {Zhang}}, \bibinfo
  {author} {\bibfnamefont {L.}~\bibnamefont {Sun}}, \bibinfo {author}
  {\bibfnamefont {B.}~\bibnamefont {Miao}}, \bibinfo {author} {\bibfnamefont
  {X.~R.}\ \bibnamefont {Wang}}, \ and\ \bibinfo {author} {\bibfnamefont
  {H.~F.}\ \bibnamefont {Ding}},\ }\href {\doibase 10.1063/5.0038567}
  {\bibfield  {journal} {\bibinfo  {journal} {Applied Physics Letters}\
  }\textbf {\bibinfo {volume} {118}} (\bibinfo {year} {2021}),\
  10.1063/5.0038567}\BibitemShut {NoStop}%
\bibitem [{\citenamefont {Zhang}\ \emph {et~al.}(2008)\citenamefont {Zhang},
  \citenamefont {Zhou}, \citenamefont {Lin}, \citenamefont {Chen},\ and\
  \citenamefont {Liaw}}]{Zhang2008}%
  \BibitemOpen
  \bibfield  {author} {\bibinfo {author} {\bibfnamefont {Y.}~\bibnamefont
  {Zhang}}, \bibinfo {author} {\bibfnamefont {Y.}~\bibnamefont {Zhou}},
  \bibinfo {author} {\bibfnamefont {J.}~\bibnamefont {Lin}}, \bibinfo {author}
  {\bibfnamefont {G.}~\bibnamefont {Chen}}, \ and\ \bibinfo {author}
  {\bibfnamefont {P.}~\bibnamefont {Liaw}},\ }\href {\doibase
  10.1002/adem.200700240} {\bibfield  {journal} {\bibinfo  {journal} {Advanced
  Engineering Materials}\ }\textbf {\bibinfo {volume} {10}},\ \bibinfo {pages}
  {534} (\bibinfo {year} {2008})}\BibitemShut {NoStop}%
\bibitem [{\citenamefont {Takeuchi}\ and\ \citenamefont
  {Inoue}(2005)}]{Takeuchi2005}%
  \BibitemOpen
  \bibfield  {author} {\bibinfo {author} {\bibfnamefont {A.}~\bibnamefont
  {Takeuchi}}\ and\ \bibinfo {author} {\bibfnamefont {A.}~\bibnamefont
  {Inoue}},\ }\href {\doibase 10.2320/matertrans.46.2817} {\bibfield  {journal}
  {\bibinfo  {journal} {MATERIALS TRANSACTIONS}\ }\textbf {\bibinfo {volume}
  {46}},\ \bibinfo {pages} {2817} (\bibinfo {year} {2005})}\BibitemShut
  {NoStop}%
\end{thebibliography}%

\end{document}